\documentclass[sigconf,authorversion,nonacm]{acmart}
\acmConference[ESEC/FSE 2023]{The 31st ACM Joint European Software Engineering Conference and Symposium on the Foundations of Software Engineering}{11 - 17 November, 2023}{San Francisco, USA}

\usepackage[utf8]{inputenc}
\usepackage{xspace}
\usepackage{booktabs}
\usepackage{hyperref}
\usepackage{xspace}
\usepackage{listings}
\usepackage{siunitx}
\usepackage{tabu}
\usepackage{tabularx}
\usepackage{caption}
\usepackage{subcaption}
\usepackage[inline]{enumitem}

\def\tabHeader{\textbf}

\newcommand*{\ie}{i.e.\@\xspace}
\makeatletter
\newcommand*{\etc}{%
    \@ifnextchar{.}%
        {etc}%
        {etc.\@\xspace}%
}
\makeatother

\definecolor{diffstart}{named}{gray}
\definecolor{diffincl}{named}{blue}
\definecolor{diffrem}{named}{red}
\lstdefinelanguage{diff}{
    basicstyle=\ttfamily\small,
    morecomment=[f][\color{diffstart}]{@@},
    morecomment=[f][\color{diffincl}]{+},
    morecomment=[f][\color{diffrem}]{-},
}
\newcommand{\toolname}{\textsc{MuRS}\xspace}
\newcommand{\toolnameDef}{\textbf{Mu}tant \textbf{R}anking \& \textbf{S}uppression\xspace}

\def\rqOne{Does \toolname decrease the overall negative feedback rate?\xspace}
\def\rqTwo{Is \toolname's ranking associated with developer feedback?\xspace}
\def\rqThree{Do \toolname's templates agree with suppression rules developed by domain experts?\xspace}

\newcommand{\PerceivedUsefulness}{\textsl{Perceived feedback}\xspace}
\newcommand{\HeuristicUsefulness}{\textsl{Killed status}\xspace}

\def\attr{\textsl}
\def\CL{\attr{CL}\xspace}
\def\Filename{\attr{Filename}\xspace}
\def\Diff{\attr{Diff}\xspace}
\def\PosFeedList{\attr{PosFeedList}\xspace}
\def\NegFeedList{\attr{NegFeedList}\xspace}
\def\KilledList{\attr{KilledList}\xspace}

\newcommand{\GoogleCommits}{\num{40000}\xspace}

\newcommand{\ExperimentGroupNegativeRate}{\num{11.45}\%\xspace}
\newcommand{\ControlGroupNegativeRate}{\num{12.41}\%\xspace}
\newcommand{\RQonePvalue}{\num{0.0242}}
\newcommand{\ABExperimentSurfacedMutants}{\num{666143}\xspace}
\newcommand{\ABExperimentsCLs}{\num{84827}\xspace}
\newcommand{\ControlGroupSurfacedMutants}{\num{365542}\xspace}
\newcommand{\ControlGroupCLs}{\num{45210}\xspace}
\newcommand{\ExperimentGroupSurfacedMutants}{\num{300601}\xspace}
\newcommand{\ExperimentGroupCLs}{\num{39618}\xspace}
\newcommand{\RQtwoExperimentRankCorrelation}{\num{0.2615}\xspace}
\newcommand{\RQtwoControlRankCorrelation}{\num{0.2658}\xspace}
\newcommand{\PvalueThreshold}{\num{0.05}\xspace}
\newcommand{\MutantsWithFeedback}{\num{23766}\xspace}
\newcommand{\ExperimentGroupWithFeedback}{\num{10941}\xspace}
\newcommand{\ExperimentGroupWithPosFeedback}{\num{9688}\xspace}
\newcommand{\ExperimentGroupWithNegFeedback}{\num{1253}\xspace}
\newcommand{\ControlGroupWithFeedback}{\num{12825}\xspace}
\newcommand{\ControlGroupWithPosFeedback}{\num{11233}\xspace}
\newcommand{\ControlGroupWithNegFeedback}{\num{1592}\xspace}
\newcommand{\MutantsWithPosFeedback}{\num{20921}\xspace}
\newcommand{\MutantsWithNegFeedback}{\num{2845}\xspace}
\newcommand{\PercentageNegMutantsSuppressed}{\num{49}\%\xspace}
\newcommand{\PercentagePosMutantsSuppressed}{\num{12}\%\xspace}
\newcommand{\NegFeedbackSuppressed}{\num{772.4159}\xspace}

\newcommand{\PosFeedbackSuppressed}{\num{1338.1659}\xspace}

\newcommand{\PercentageNotUsefulTopHalf}{\num{16.39}\%\xspace}
\newcommand{\NumNotUsefulTopHalf}{\num{261}\xspace}

\title[\toolname: Mutant Ranking and Suppression using Identifier Templates]{\toolname: Mutant Ranking and Suppression\\using Identifier Templates}

\author{Zimin Chen}
\authornote{{\label{authornote1}}Work done at Google.}
\affiliation{%
  \institution{KTH Royal Institute of Technology}
  \city{Stockholm}
  \country{Sweden}
}
\email{zimin@kth.se}

\author{Malgorzata Salawa}
\affiliation{%
  \institution{Google}
  \city{Zurich}
  \country{Switzerland}}
\email{magorzata@google.com}

\author{Manushree Vijayvergiya}
\affiliation{%
  \institution{Google}
  \city{Zurich}
  \country{Switzerland}}
\email{manushree@google.com}

\author{Goran Petrovic}
\affiliation{%
  \institution{Google}
  \city{Zurich}
  \country{Switzerland}}
\email{goranpetrovic@google.com}

\author{Marko Ivanković}
\affiliation{%
  \institution{Google}
  \city{Zurich}
  \country{Switzerland}}
\email{markoi@google.com}

\author{Ren{\'e} Just\textsuperscript{\ref{authornote1}}}
\affiliation{%
  \institution{University of Washington}
  \city{Seattle}
  \country{USA}}
\email{rjust@cs.washington.edu}

\begin{document}

\begin{abstract}
Diff-based mutation testing is a mutation testing approach that only mutates lines affected by a code change under review.  This approach scales to a code base of billions of lines of code and tens of thousand commits per day. It introduces test goals (mutants) that are directly relevant to an engineer's goal such as fixing a bug, adding a new feature, or refactoring existing functionality.
Google's mutation testing service integrates diff-based mutation testing into the code review process and continuously gathers developer feedback on mutants surfaced during code review. To enhance the developer experience, the mutation testing service implements a number of suppression rules, which target not-useful mutants---that is, mutants that have consistently received negative developer feedback.
However, while effective, manually implementing suppression rules require significant engineering time. An automatic system to rank and suppress mutants would facilitate the maintenance of the mutation testing service.

This paper proposes and evaluates \toolname, an automated approach that groups mutants by patterns in the source code under test and uses these patterns to rank and suppress future mutants based on historical developer feedback on mutants in the same group.
To evaluate \toolname, we conducted an A/B testing study, comparing \toolname to the existing mutation testing service.
Despite the strong baseline, which uses manually developed suppression rules, the results show a statistically significantly lower negative feedback ratio of \ExperimentGroupNegativeRate for \toolname versus \ControlGroupNegativeRate for the baseline. The results also show that \toolname is able to recover existing suppression rules implemented in the baseline. Finally, the results show that statement-deletion mutant groups received both the most positive and negative developer feedback, suggesting a need for additional context that can distinguish between useful and not-useful mutants in these groups.
Overall, \toolname has the potential to substantially reduce the development and maintenance cost for an effective mutation testing service by automatically learning suppression rules.
\end{abstract}


    
\maketitle

\section{Introduction}
\label{sec:intro}

Software testing is an essential part of software development that validates and verifies the software under test. A well tested program is an indication of a reliable program~\cite{kochhar2015code}. But how do we assess whether a program is well tested? One way is to use mutation testing~\cite{acree1979mutation}, which systematically introduces small changes into the program and checks whether the program's tests detect the changes. Mutation testing has been widely studied~\cite{jia2010analysis} and sees increasing adoption in industry~\cite{beller2021would, petrovic2021practical}. 

Mutation testing systematically applies mutation operators that create program variants called mutants. Each mutant differs from the original program by a small change such as a changed literal, a changed operator, or a deleted statement. Even a small number of mutation operators, when systematically applied to an entire program, can generate vast numbers of mutants, and evaluating all of them is computationally expensive~\cite{wong1995reducing}---prohibitively so for an industry-level code base. To solve this problem, \textit{diff-based mutation testing}~\cite{petrovic2021practical} incrementally mutates, during code review, only source code lines that are affected by the code change under review.
    
\begin{figure}
\centering
\includegraphics[width=\linewidth,keepaspectratio]{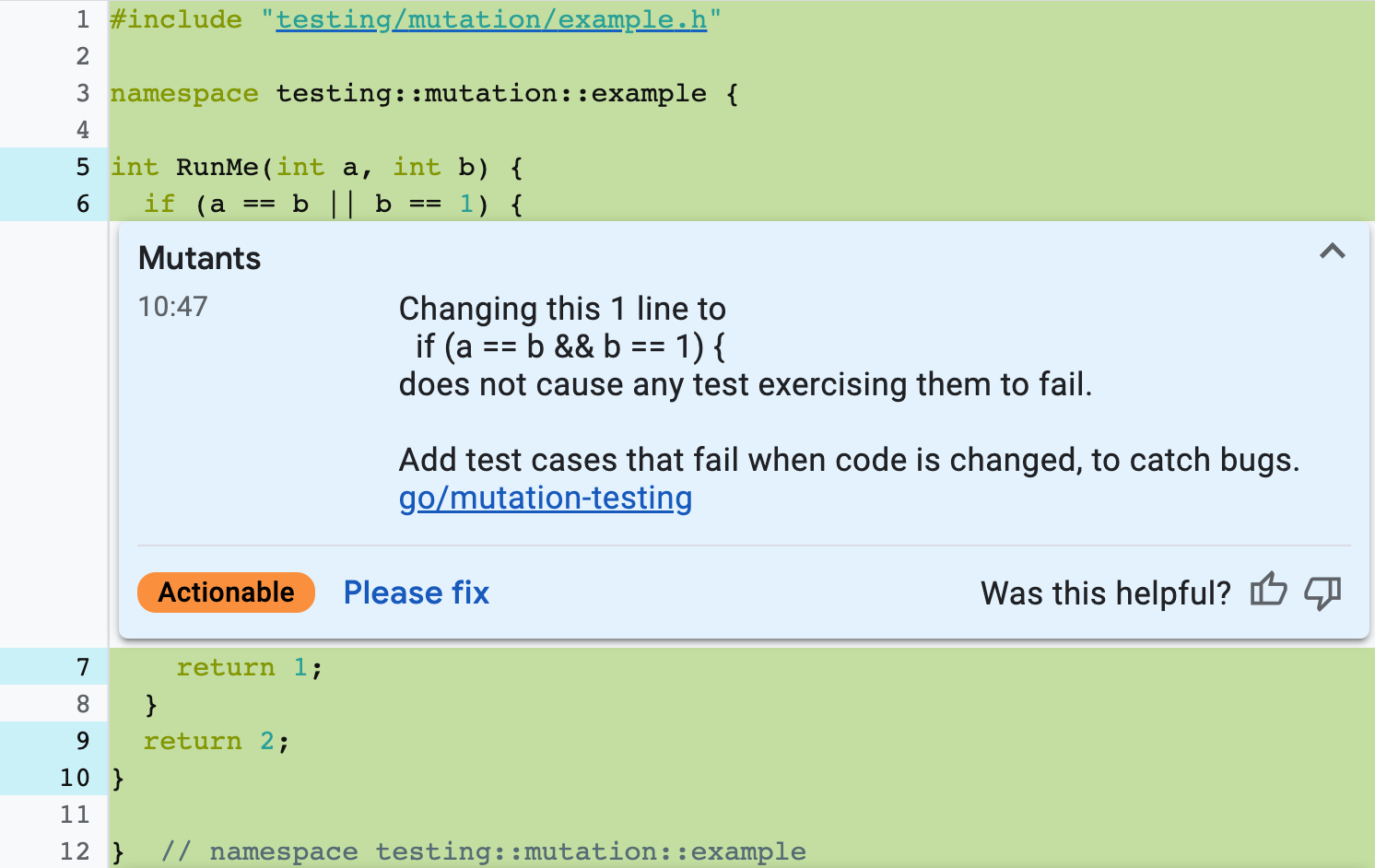}
\caption{A mutant surfaced in the code review tool. A reviewer can click ``Please fix'', indicating that the author should resolve the mutant. Both author and reviewers can click ``Thumbs up'' and ``Thumbs down'' to give feedback.}
\label{fig:mutant_in_critique}
\end{figure}

Google has successfully deployed diff-based mutation testing for a code base that sees more than \GoogleCommits code commits per day to more than a billion lines of code~\cite{petrovic2021practical}. The corresponding code-review tool offers an option for authors and reviewers to provide feedback for mutants surfaced during code review (see \autoref{fig:mutant_in_critique}). If a reviewer clicks the ``Please fix'' button, the code-review tool auto-generates a review comment, asking the author to resolve the mutant. Additionally, both authors and reviewers can click ``Thumbs up'' or ``Thumbs down'' to give positive or negative feedback to the surfaced mutants. The negative feedback ratio is regularly measured and serves as an indicator for effective false positives~\cite{sadowski2018lessons}. To lower that ratio, the existing mutation testing service implements a number of static suppression rules that prevent mutants that are likely to received negative feedback from being surfaced, and it prioritizes mutation operators based on the historical effectiveness of the mutants they generate. Over the years, the manually developed suppression and prioritization rules have reduced the negative feedback ratio from over 80\% to well below 15\%~\cite{petrovic2021practical}.
However, manually implementing rules is labor-intensive, the rules require regular maintenance, and prioritization based on mutation operators may be too coarse-grained.

This paper presents \toolname (\toolnameDef) that aims to tackle these problems and further reduce the negative feedback ratio. \toolname uses identifier templates---an abstraction over the unified diff between the mutant and the original program. Specifically, \toolname takes, for a given mutant, the unified diff and replaces (1) literals with their type names (e.g., int or String) and (2) identifiers with {\small\texttt{IDENTIFIER}}. \toolname optionally preserves the most common literals and identifiers, using a configurable vocabulary size. Similarly, the number of context lines in the unified diff considered when building identifier templates is another hyper-parameter.
\toolname then groups all historical mutants and aggregates their developer feedback by identifier template. The aggregated feedback for each identifier template is then used to rank and suppress newly generated mutants. The intuition behind this approach is that mutants whose identifier templates have largely received positive feedback should be ranked highly, whereas mutants whose identifier templates have received negative feedback should be ranked lower or suppressed altogether. \toolname scales very well to a large code base: our analyses computed millions of identifier templates on a machine with commodity hardware and only 16 GB of RAM.

We conducted an A/B testing study to evaluate \toolname' performance, randomly assigning 50\% of code reviews at Google during the evaluation period to the experiment group (\toolname) and all other code reviews to a control group (existing mutation testing service). During the evaluation period, a total of \ABExperimentSurfacedMutants mutants surfaced across the five languages supported by \toolname (Python, Java, C++, Go, and TypeScript). The key results are as follows:

\begin{itemize}
\item The overall negative feedback ratio for the experiment group is \ExperimentGroupNegativeRate, which is statistically significantly lower than the negative feedback ratio of \ControlGroupNegativeRate for the control group.

\item The correlation coefficient (Kendall's Tau-b) between \toolname's ranking score and the ranked mutants' perceived usefulness (developer feedback) is \RQtwoExperimentRankCorrelation, which indicates a moderate correlation. Further, the top 50\% of ranked mutants only contain \PercentageNotUsefulTopHalf of mutants with negative feedback. This suggests that \toolname's ranking of mutants effectively decreases the likelihood of surfacing not-useful mutants.

\item Choosing a suppression threshold involves a trade-off: when applied to all surfaced mutants in the control group, \toolname would have suppressed \PercentageNegMutantsSuppressed of mutants that received negative feedback, and it would have suppressed \PercentagePosMutantsSuppressed of mutants that received positive feedback. The trade-off between correctly suppressing not-useful mutants, which waste engineers' time, and incorrectly suppressing useful mutants, which likely lead to additional tests, is a trade-off that one has to consider when deploying \toolname.

\item For all supported languages, the identifier templates that receive the most controversial feedback (i.e., the most positive and the most negative feedback) are all statement removal mutants. Prior research has found that statement removal mutants are coupled with real faults more often than other types of mutants \cite{just2014mutants}. This finding suggests that although statement removal mutants are useful in general, further investigation is necessary to identify the specific subset of statement removal mutants that are not useful.

\end{itemize}

We additionally performed a retrospective analysis, investigating what identifier templates \toolname would have learned and scored at different points in the past six years. A manual analysis of the most common identifier templates that consistently received negative feedback showed that \toolname is able to recover suppression rules implemented in the existing mutation testing service.

\section{Terminology}
\label{sec:terminology}

This section defines the terminology used throughout the paper:
\begin{itemize}
\item \textbf{Changelist}: A set of changes made to the code base, which are submitted for review (similar to a merge request).
\item \textbf{Snapshot}: A specific point in time, capturing the state of a changelist as it evolves during the code review process. Each changelist has one or more snapshots.
\item \textbf{Generated mutants}: All mutants generated by the mutation testing service.
\item \textbf{Surviving mutants}: Mutants that are not detected by any of the existing tests---a subset of all generated mutants.
\item \textbf{Surfaced mutants}: Mutants displayed in the code review tool---a subset of surviving mutants. The number of surfaced mutants is limited to avoid overwhelming developers.
\item \textbf{Positive feedback}: A surfaced mutant receives positive feedback when the ``Please Fix'' or ``Thumbs Up'' button is clicked in \autoref{fig:mutant_in_critique}; we deem such mutants useful.
\item \textbf{Negative feedback}: A surfaced mutant receives negative feedback when the ``Thumbs Down'' button is clicked in \autoref{fig:mutant_in_critique}; we deem such mutants not useful.
\end{itemize}

\begin{figure*}
    \centering
    \includegraphics[width=0.95\textwidth, keepaspectratio, trim={25mm 0 0 0},clip]{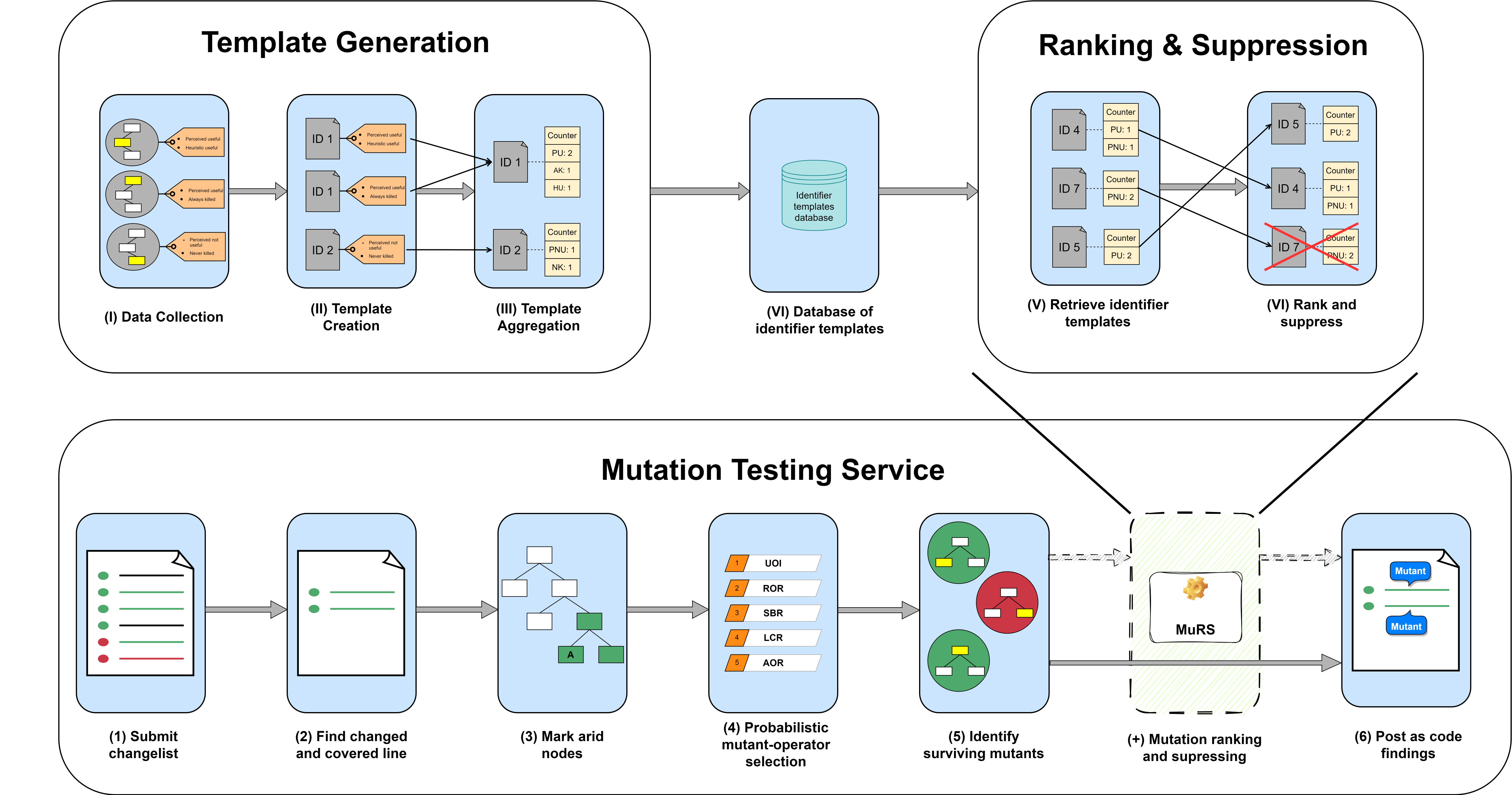}
    \caption{Mutation Testing Service (bottom) shows the end-to-end workflow from submitting a changelist for code review to surfacing mutants as code findings in the code review tool: (1) A changelist is submitted for review. Red and green dots indicate that a line is (not) covered by existing tests. A green, red, or black line indicates that it is added, removed, or unchanged, respectively. (2) Lines that are both changed and covered are identified. (3) Arid nodes, which are not eligible for mutation, are marked with `A'. (4) Mutation operators are applied in order to generate at most one mutant per changed line. (5) Test cases are run and surviving mutants are identified. (6) A random subset of surviving mutants are surfaced as code findings.
    \toolname, is added between (5) and (6): it ranks and suppresses surviving mutants as opposed to randomly sampling from all surviving mutants. \toolname consists of two phases, \textbf{Template Generation} and \textbf{Ranking \& Suppression}. \textbf{Template Generation} consists of three stages: (I) Historical mutants and their feedback are collected. (II) Identifier templates are generated for each mutant. (III) Mutants are grouped by identifier template and feedback scores are aggregated. (IV) The identifier templates are stored in a database. \textbf{Ranking \& Suppression} consists of two stages: (V) The aggregated feedback score is retrieved from the database for each surviving mutant's identifier template. (VI) Surviving mutants are ranked and suppressed based on their feedback score.}
    \label{fig:flowchart}
\end{figure*}
\section{Mutation Testing at Google}
\label{sec:background}

Integrating mutation testing into a large code base, like Google's, poses significant challenges related to scalability and workflow integration. Generating and testing all possible mutants is computationally infeasible, and reporting all surviving mutants to developers would be overwhelming. To overcome these challenges, Google's mutation testing service implements \textit{diff-based mutation testing}: it only generates mutants for the lines of code that have been changed in a given changelist. It also implements \textit{arid node detection} and \textit{probabilistic mutation-operator selection} to surface mutants that developers mostly consider useful in practice~\cite{petrovic2021practical}.

In \autoref{fig:flowchart}, ``Mutation Testing Service'' illustrates the entire process, from submitting a changelist for code review to surfacing mutants as findings in the code review tool. For a more comprehensive view of mutation testing at Google, we refer the reader to \cite{petrovic2021practical}. This section focuses on a higher-level description and only details the steps that are relevant to our approach:

\begin{enumerate}
\item A developer submits a changelist for code review.
\item The mutation testing service identifies all changed lines that are covered by at least one test.
\item The mutation testing service builds the abstract syntax tree of each affected file and marks arid nodes that are not eligible for mutation. Mutating arid nodes produces non-actionable mutants, such as String mutations in logging statements, for which developers justifiably would not write tests.~\label{enu:arid_nodes}
\item The mutation testing service uses \textit{probabilistic mutation-operator selection} that prioritizes mutation operators based on their historical performance until at most one mutant is generated for each affected line. The historical performance is computed for each mutation operator and programming language, which is relatively coarse grained.~\label{enu:mutation_selection}
\item The mutation testing service identifies surviving mutants by executing the existing tests against each generated mutant.~\label{enu:surviving_mutants}
\item The mutation testing service surfaces a random sample of surviving mutants in the code review tool, based on per file and per changelist thresholds.~\label{enu:code_findings}
\end{enumerate}

\toolname aims to address two limitations of the current mutation testing service, related to the implementation of arid nodes (step \ref{enu:arid_nodes}) and the probabilistic mutation-operator selection (step \ref{enu:mutation_selection}). While arid-node detection is effective, it currently requires human effort to identify and implement a static rule for suppressing a certain group of mutants. This implementation is specific to each programming language and must be manually maintained as the code base evolves. As for probabilistic mutation-operator selection, the problem is that the selection is coarse-grained and context-agnostic. Currently, the mutation testing service implements five mutation operators: AOR (Arithmetic operator replacement), LCR (Logical connector replacement), ROR (Relational operator replacement), UOI (Unary operator insertion), and SBR (Statement block removal). However each mutation operator corresponds to a number of possible mutations, such as replacing addition with subtraction, division, or multiplication. In contrast, \toolname considers the actual mutation and the code before and after as additional context.

\section{\toolname: \toolnameDef}
\label{sec:approach}

A key goal of \toolname is to reduce the negative feedback ratio of surfaced mutants. The negative feedback ratio is defined as the number of surfaced mutants with negative feedback divided by the total number of surfaced mutants with positive or negative feedback.
\toolname implements a template-based approach for five languages (C++, Java, Go, Python and TypeScript) that ranks and suppresses mutants based on historical developer feedback. It consist of two phases, as shown in the upper part of \autoref{fig:flowchart}:
\begin{enumerate}
    \item \textbf{Template Generation}: During this phase, \toolname generates identifier templates from all past mutants, associating each identifier template with a list of metrics such as number and type of received feedback, and number of times it was generated and killed. This phase is intended to run periodically to collect identifier templates with all statistics until that point of time and store them in a database. The time interval for template re-computation can be chosen based on the needs and resource constraints of the users, balancing the importance of the data freshness with the acceptable resource consumption. In our case, this phase run every two months.
     \item \textbf{Ranking \& Suppression}:  During this phase, \toolname uses the latest identifier templates stored in a database to rank and suppress mutants in production. Specifically, it uses the list of metrics to represent the past performance associated with each identifier template to rank and suppress surviving mutants that match an identifier template.
\end{enumerate}

There are different ways to generate the identifier templates and to rank and suppress mutants. Sections~\ref{sec:approach:templates} and~\ref{sec:approach:rankingsuppressing} discuss these configuration options in detail, and section~\ref{sec:approach:find_best_settings} describes the configuration that we chose based on a preliminary evaluation. All of \toolname' identifier templates can be stored in a lookup table in memory, using less than 16GB of RAM.

\begin{table}[t]
\renewcommand{\arraystretch}{1.2} 
\begin{tabularx}{\linewidth}{lX}
\toprule
\tabHeader{Attribute} & \tabHeader{Explanation} \\
\midrule
\CL &  The changelist ID \\
\Filename &  The name of the mutated file \\
\Diff & The unified diff with all surrounding lines as context between the original file and the mutant     \\
\PosFeedList & A list of type boolean (one per snapshot) indicating whether a mutant received a positive feedback at that snapshot \\
\NegFeedList & A list of type boolean (one per snapshot) indicating whether a mutant received a negative feedback at that snapshot \\
\KilledList & A list of type boolean (one per snapshot) indicating whether a mutant was killed at that snapshot \\
\bottomrule
\end{tabularx}
\caption{The attributes collected for each mutant in the \textit{Data Collection} stage.}
\label{tab:mutant_attributes}
\renewcommand{\arraystretch}{1.0}
\end{table}

\subsection{Template Generation}
\label{sec:approach:templates}

Template Generation consists of three stages, as shown in \autoref{fig:flowchart}:
\begin{itemize}
    \item[(I)] \textbf{Data Collection}: Gather all previously generated mutants along with their associated attributes (see \autoref{tab:mutant_attributes}).
    \item[(II)] \textbf{Template Creation}: Create identifier templates for all previously generated mutants.
    \item[(III)] \textbf{Template Aggregation}: Aggregate mutants with the same identifier templates.
\end{itemize}

\subsubsection{Data Collection}

In this stage, \toolname gathers mutants and their attributes from a database of all previously generated mutants. \toolname then collects various attributes (\autoref{tab:mutant_attributes}) for each mutant and assigns two kinds of labels to each mutant, based on said attributes: \PerceivedUsefulness and \HeuristicUsefulness.

\smallskip
\noindent
\textbf{\PerceivedUsefulness} represents a mutant's usefulness as perceived by developers. It is derived from \PosFeedList and \NegFeedList:
\begin{itemize}
\item \textbf{Perceived useful}: \PosFeedList contains at least one \textit{True} and \NegFeedList is all \textit{False}.
\item \textbf{Perceived not-useful}: \PosFeedList is all \textit{False} and \NegFeedList contains at least one \textit{True}.
\item \textbf{Mixed feedback}: Both \PosFeedList and \NegFeedList contain at least one \textit{True}.
\item \textbf{No feedback}: Both \PosFeedList and \NegFeedList only contain \textit{False}.
\end{itemize}

\smallskip
\noindent
\textbf{\HeuristicUsefulness} is a label representing whether a mutant was killed (or not) throughout the code review. It is derived from \KilledList:
\begin{itemize}
\item \textbf{Always killed}: All entries in \KilledList are \textit{True}.
\item \textbf{Never killed}: All entries in \KilledList are \textit{False}.
\item \textbf{Eventually killed}: The \KilledList is a sequence of \textit{False}, followed by a sequence of \textit{True}, which means that a mutant survived up to a certain point in the code review process and was consistently killed afterwards.
\item \textbf{Mixed killed}: None of the above conditions are met. This can happen when the entries in \KilledList alternate between \textit{False} and \textit{True}.
\end{itemize}

At the end of this stage, each mutant is assigned an appropriate \PerceivedUsefulness and \HeuristicUsefulness label.

\subsubsection{Template Creation}

In this stage, \toolname generates identifier templates for all labeled mutants based on the \Diff attribute.

We define an abstraction over the unified diff between the mutated and the original code at the line level, which we refer to as an \textit{identifier template}. The abstraction removes comments from the code before considering the diff, including context lines (where applicable). We chose this abstraction because it is relatively fast to compute and allows for quick inference, even in the presence of millions of mutants. To generate identifier templates, we can control the template's specificity using different abstraction levels and various parameters, which we subsequently explain using the mutant in \autoref{lst:mutant_example} as a running example.

\begin{figure}[t]
\begin{lstlisting}[language=diff,columns=flexible, frame=tb, basicstyle=\small, label={lst:mutant_example}, caption={An example of a mutant in Python.}, captionpos=b, breaklines=true]

 def is_even(x):
- return x %
+ return x %

 def is_odd(x):
\end{lstlisting}
\end{figure}

\smallskip
\noindent
\textbf{Abstraction levels} that we explored are:
\begin{itemize}
\item \textbf{Original code template}: Use each changed line as is.
\begin{lstlisting}[language=diff,columns=flexible, frame=tb, basicstyle=\small, label={lst:original_code_template}, breaklines=true]
- return x %
+ return x %
\end{lstlisting}

\item \textbf{Typed template}: Replace all literals with their corresponding type name, and identifiers with \textit{IDENTIFIER}. For instance, integer literals are replaced with \textit{INT}. However, all keywords such as \textit{while}, \textit{for} and \textit{return} are kept.
\begin{lstlisting}[language=diff,columns=flexible, frame=tb, basicstyle=\small, label={lst:typed_template}, breaklines=true]
- return IDENTIFIER %
+ return IDENTIFIER %
\end{lstlisting}

\item \textbf{Indexed typed template}: Similar to Typed template, but index each type name. This means that each name is followed by a number, in order of appearance, to differentiate between identifiers of the same type. For example, the two integers, $2$ and $0$, are replaced with INT\_0 and INT\_1, respectively.
\begin{lstlisting}[language=diff,columns=flexible, frame=tb, basicstyle=\small, label={lst:indexed_typed_template}, captionpos=b, breaklines=true]
- return IDENTIFIER %
+ return IDENTIFIER %
\end{lstlisting}

\end{itemize}

\smallskip
\noindent
\textbf{Identifier template parameters} that we explored are:
\begin{itemize}
\item \textbf{Context size}: The number of lines preceding and succeeding the mutated line(s).
Larger context sizes lead to more specificity, because more context lines are included and the corresponding identifier template matches fewer, but more specific types of mutants. The following identifier template corresponds to an Indexed typed template with a context size of 1.
\begin{lstlisting}[language=diff,columns=flexible, frame=tb, basicstyle=\small, label={lst:indexed_typed_template_with_context}, breaklines=true]
 def IDENTIFIER_0(IDENTIFIER_1):
- return IDENTIFIER_1 %
+ return IDENTIFIER_1 %

\end{lstlisting}

\item \textbf{Vocabulary size}: The number of the most common variable names and literals to keep without replacement. Larger vocabulary sizes lead to more specificity. For example, if \textit{numpy} is part of the vocabulary, mutants in the context of the identifier \texttt{numpy} can be ranked and suppressed differently than similar mutants in other contexts. The following identifier template corresponds to an Indexed typed template with a vocabulary containing \textit{x}, but not containing $2$ and $0$.
\begin{lstlisting}[language=diff,columns=flexible, frame=tb, basicstyle=\small, label={lst:indexed_typed_template_with_vocab}, breaklines=true]
- return x %
+ return x %
\end{lstlisting}

\end{itemize}

\subsubsection{Template Aggregation}

Similar mutants will generate the same identifier template depending on the abstraction level and identifier template parameters. In this stage, \toolname aggregates all labeled identifier template instances---counting the number of \PerceivedUsefulness and \HeuristicUsefulness labels. PU, PNU, MF and NF counts perceived useful, perceived not-useful, mixed feedback and no feedback of the \PerceivedUsefulness label. AK, NK, EK and MK counts always killed, never killed, eventually killed and mixed killed of the \HeuristicUsefulness label.

All identifiers with their corresponding counters are then stored in a database, shown in (IV) of \textbf{Template Generation} in \autoref{fig:flowchart}. These identifier templates are then used to populate a lookup table and efficiently rank and suppress mutants in the Ranking \& Suppression phase.

\subsection{Ranking \& Suppression}
\label{sec:approach:rankingsuppressing}

Ranking \& Suppression consumes the pre-computed identifier templates and their associated counters, as derived from the Template Generation phase. The intuition is that we can use these counters to rank and suppress mutants. For example, identifier templates with high PU counts may be considered more useful than those with low PU counts, hence should be ranked higher. While identifier templates with high PNU counts may be considered highly not useful and hence, can be suppressed. Ranking and suppression are two independent procedures, as the former only gives a relative ordering between mutants, so the highest ranked mutants may still not be worth surfacing to the developer. The following sections describe how we rank and suppress mutants.

\subsubsection{Ranking}

\toolname ranks identifier templates and matching mutants based on a usefulness score, which comprises two distinct scores: (1) the developer-feedback score, derived from the \PerceivedUsefulness label, and (2) the killed score, derived from the \HeuristicUsefulness label. \toolname ranks the identifier templates based on the developer-feedback score, and breaks ties based on the killed score.

We explored two definitions for the developer-feedback score:
\begin{itemize}
    \item \textbf{Usefulness score}: This score is the ratio of PU over PU+PNU. It is calculated as $\frac{PU}{PU+PNU}$.
    \item \textbf{Bayes usefulness score}: It uses the \textbf{Usefulness score} and the average \textbf{Usefulness score} for all identifier templates to calculate a bayesian weighted average. Let $m$ be the average number of $PU+PNU$ per identifier template, and $average$ be the average \textbf{Usefulness score}. The weight $w$ is defined as $\frac{PU+PNU}{PU+PNU+m}$. Using this weight, the \textbf{Bayes usefulness score} is defined as $w * \textbf{Usefulness score} + (1 - w) * average$. Compared to \textbf{Usefulness score}, this score accounts for noise when $PU+PNU$ is low, \ie when there is not enough feedback from developers.
\end{itemize}

We also explored two definitions for killed score:
\begin{itemize}
\item \textbf{Kill-ratio score}: It is defined as $\frac{K}{G}$, representing the percentage of times that this type of mutant is killed. 
\item \textbf{Kill-counter score}: It is a tuple with (AK, EK, MK, NK) counters. The ordering between the counters is based on their correlation with the developer feedback.
\end{itemize}

Since there are 2 developer-feedback scores and 2 killed scores, there are 4 different combinations of usefulness scores.

\subsubsection{Suppression}
\label{sec:approach:suppress_mutants}

\toolname aims to suppress mutants that are likely to receive negative feedback. When designing the suppress function, we chose to implement the following two desired properties:

\begin{enumerate}
    \item If an identifier template is not in the database (\ie the mutant has not been encountered before), \toolname would not suppress it. This allows for gathering feedback for new identifier templates.
    \item Only the PU and PNU counters are considered for suppressing mutants, as it is crucial not to over-suppress mutants. Suppressing an identifier template means that all mutants with the corresponding identifier template will not be surfaced, therefore we would like to be careful and only suppress mutants based on labels derived from the developer feedback, \ie PU and PNU.
\end{enumerate}

We explored the following three suppression functions:
\begin{itemize}
    \item \textbf{No suppression}: Do not suppress any mutants. This is added to experiment if a suppression function is needed at all.
    \item \textbf{Average threshold}: Suppress mutants whose \textbf{Usefulness score} is lower than the average \textbf{Usefulness score} across all identifier templates.
    \item \textbf{Probabilistic}: Use the p-value of the z-test to suppress the mutants with a calculated probability, as the \textbf{Usefulness score} follows a normal distribution. For example if the z-score is $-1$, the corresponding p-value is $0.1587$, then we have a $1-0.1587=0.8413$ chance of suppressing this mutant. This suppress function is only applied when the \textbf{Usefulness score} is lower than the average \textbf{Usefulness score} across all identifier templates.
\end{itemize}

At the end of the Ranking \& Suppression phase, \toolname has ranked all generated mutants and suppressed mutants that are likely to receive negative feedback. At this point, the mutation testing service continues to surfacing the selected mutants as findings in the code review tool.

\subsection{Hyperparameter Tuning}
\label{sec:approach:find_best_settings}

\begin{table}[t]
\begin{tabu}{cc|cc}
\toprule
\multicolumn{2}{c|}{\tabHeader{Original}} & \multicolumn{2}{c}{\tabHeader{\toolname}}\\
\midrule
Mutant      & Label             & Mutant            & Label                \\
\cmidrule(r){1-2}
\cmidrule(l){3-4}
A           & Positive feedback & A                 & Positive feedback    \\
B           & No feedback       & B                 & No feedback          \\
C           & Negative feedback & D                 & Negative feedback    \\
D           & Negative feedback &                   &                      \\
\bottomrule
\end{tabu}
\caption{Hyper-parameter tuning for \toolname. Mutants on the left are surfaced without using \toolname. Mutants on the right are surfaced using \toolname with a specific hyper-parameter setting that suppresses mutant C. In this case, the original negative feedback rate is $\frac{2}{4} = 0.5$, and it is $\frac{1}{3} \approx 0.33$ for \toolname. The best hyper-parameter setting is the one that achieves the lowest negative feedback rate.}
\label{tab:not_useful_example}
\end{table}

\toolname's design space involves five dimensions:
(1) template type (original code template, typed template or indexed typed template),
(2) vocabulary size (0, 1000, 5000, or 10000),
(3) context size (0 or 1),
(4) ranking function (4 combinations of developer-feedback score and killed score), and
(5) suppression function (no suppression, average threshold or probabilistic).

We conducted a preliminary evaluation of all possible hyperparameter combinations using historical mutants. Specifically, we used the mutants generated before July 2022 to generate identifier templates and computed the hypothetical negative feedback ratio for each hyperparameter setting on the mutants generated in July 2022. We then selected the setting with the lowest ratio. The example in \autoref{tab:not_useful_example} illustrates how the hypothetical negative feedback ratio is calculated. On the left we have mutants with their labels, note that we have the label because they have already been generated and shown to the developers. On the right is an example of how one combination might rank and suppress the mutants, where mutant C is suppressed. In this example, the original negative feedback ratio is $\frac{2}{4} = 0.5$, the hypothetical negative feedback ratio for \toolname is $\frac{1}{3} \approx 0.33$. The best hyperparameters combination is the one that achieves the lowest hypothetical negative feedback ratio.

The hyperparameter tuning identified the following setting as the best:
(1) \textbf{Indexed type template},
(2) \textbf{vocabulary size = 0},
(3) \textbf{context size = 0}
(4) \textbf{Bayes usefulness score with kill-counter score}, and
(5) \textbf{Probabilistic suppression function}.
We consistently used this hyperparameter setting to answer our research questions in \autoref{sec:eval}.
\section{Evaluation Methodology}
\label{sec:eval}

To evaluate \toolname, we compared the current mutation testing service, which randomly selects $n$ mutants and surfaces them as code findings in the code-review tool, to an enhanced variant, which surfaces the top-$n$ mutants after ranking and suppressing mutants with \toolname (see \autoref{fig:flowchart}). Note that both variants use existing optimizations, such as arid-node suppression and probabilistic mutation-operator selection~\cite{petrovic2021practical}. Moreover, the mutation testing service limits the number of surfaced mutants during code review to avoid overwhelming developers: it surfaces at most three mutants per file and at most ten mutants per changelist. We kept these thresholds constant for both variants. 

We followed a standard A/B testing methodology for changelists subject to mutation by the mutation testing service: each changelist was randomly assigned to the experiment group with probability $p=0.5$; all other changelists formed the control group. Regardless of group, each mutant that surfaced in the code review tool was subject to developer feedback (i.e., positive feedback, negative feedback, or no feedback at all).
The A/B testing allowed us to answer the following two research questions:
\begin{itemize}
\item [\textbf{RQ1}] \rqOne
\smallskip
\item [\textbf{RQ2}] \rqTwo
\end{itemize}

The existing mutation testing service has been deployed for over seven years, with domain experts manually developing arid-node heuristics to suppress mutants that will likely receive negative feedback~\cite{petrovic2021practical}. The majority of these heuristics were implemented early during development and deployment, others were added over time in response to user feedback. Considering historical mutant data, we additionally answered the following research question:
\begin{itemize}
\item [\textbf{RQ3}] \rqThree
\end{itemize}

At a high level:
RQ1 uses an extrinsic evaluation and determines whether \toolname can improve mutant suppression over a manually tuned baseline;
RQ2 uses an intrinsic evaluation and determines the degree to which \toolname's ranking of mutants is correlated with their perceived usefulness.
RQ3 uses case studies and determines whether \toolname's templates and their statistics can be linked to manually developed suppression rules.

\subsection{RQ1: \rqOne}
\label{sec:eval:rq1}
During the A/B testing evaluation period\footnote{Due to industry confidentiality reasons, we cannot disclose the exact time range, but it was a span of multiple weeks starting in July 2022.}, \ABExperimentSurfacedMutants mutants surfaced across \ABExperimentsCLs changelists.
For the control group, \ControlGroupSurfacedMutants mutants surfaced for \ControlGroupCLs changelists.
For the experiment group, \ExperimentGroupSurfacedMutants mutants surfaced for \ExperimentGroupCLs changelists.
For each group we computed the overall negative feedback ratio and compared the results. Additionally, we conducted a two-sample hypothesis test. Specifically, we used the chi-square test of independence~\cite{pearson1900x}, testing the null hypothesis that developer feedback is independent of group association (\ie, experiment vs. control group). We set the significance threshold to \PvalueThreshold.

\subsection{RQ2: \rqTwo}
\label{sec:eval:rq2}
The mutation testing service may generated more surviving mutants than can be surfaced in the code-review tool. Thus, it is important to rank mutants that are likely to receive positive feedback higher. To determine to what extent \toolname succeeds in properly ranking mutants, we computed the rank of all surfaced mutants in the control group and correlated that rank with the mutants' received feedback. Specifically, we computed Kendall's Tau-b (accounting for expected ties in the dichotomous feedback variable) between \toolname's ranking score and developer feedback. We again set the significance threshold to \PvalueThreshold. We repeated the same analysis for all mutants in the experiment group.
Additionally, we computed the ratios of mutants that already received positive or negative feedback that would have been suppressed in the control group, investigating the trade-off between suppressing not-useful mutants while retaining useful ones.

\subsection{RQ3: \rqThree}
\label{sec:eval:rq3}
We wished to better understand whether \toolname's identifier templates are linked to arid nodes, for which the existing mutation testing service suppresses all mutants. However, not-useful mutants can no longer be observed once a suppression rule is implemented. Therefore, we resorted to historical mutant data to determine whether \toolname, if applied to not-useful mutants prior to suppression, would have identified said mutants. Specifically, We compiled a data set with all surfaced mutants generated since 2017. For each mutant, we computed the corresponding identifier template and aggregated template data per month. We then performed a case study on a sample of templates that produced many mutants, consistently rated as not useful.
\section{Evaluation Results}
\label{sec:result}

\subsection{RQ1: \rqOne}

\begin{table}[t]
\begin{tabular}{lll|l}
\toprule
                 & \tabHeader{Positive}         & \tabHeader{Negative}    & \tabHeader{Total} \\
\midrule
\tabHeader{Experiment group} & \ExperimentGroupWithPosFeedback           & \ExperimentGroupWithNegFeedback          & \ExperimentGroupWithFeedback  \\[4pt]
\tabHeader{Control group}    & \ControlGroupWithPosFeedback          & \ControlGroupWithNegFeedback          & \ControlGroupWithFeedback  \\
\midrule
\textbf{Total}  & \MutantsWithPosFeedback & \MutantsWithNegFeedback & \MutantsWithFeedback  \\
\bottomrule
\end{tabular}
\caption{Mutants with feedback during A/B testing.}
\label{tab:contigency_table}
\vspace*{-8pt}
\end{table}

During the A/B testing evaluation period, a total of \MutantsWithFeedback mutants received developer feedback. \autoref{tab:contigency_table} shows the distribution among the two groups and positive vs. negative feedback.
The negative feedback ratio for mutants in the experiment group is \ExperimentGroupNegativeRate (\ExperimentGroupWithNegFeedback out of \ExperimentGroupWithFeedback mutants).
In contrast, the negative feedback ratio for mutants in the control group is \ControlGroupNegativeRate (\ControlGroupWithNegFeedback out of \ControlGroupWithFeedback mutants).
The Chi-squared test of independence for the two variables in \autoref{tab:contigency_table} yields a p-value of \RQonePvalue. Since this value is lower than our predefined significance threshold of \PvalueThreshold, we can reject the null hypothesis. This means that the observed differences in feedback are not independent of the group---that is, the lower negative feedback ratio of \ExperimentGroupNegativeRate in the experiment group is statistically significant.

\subsection{RQ2: \rqTwo}
\begin{figure}
\includegraphics[width=\linewidth]{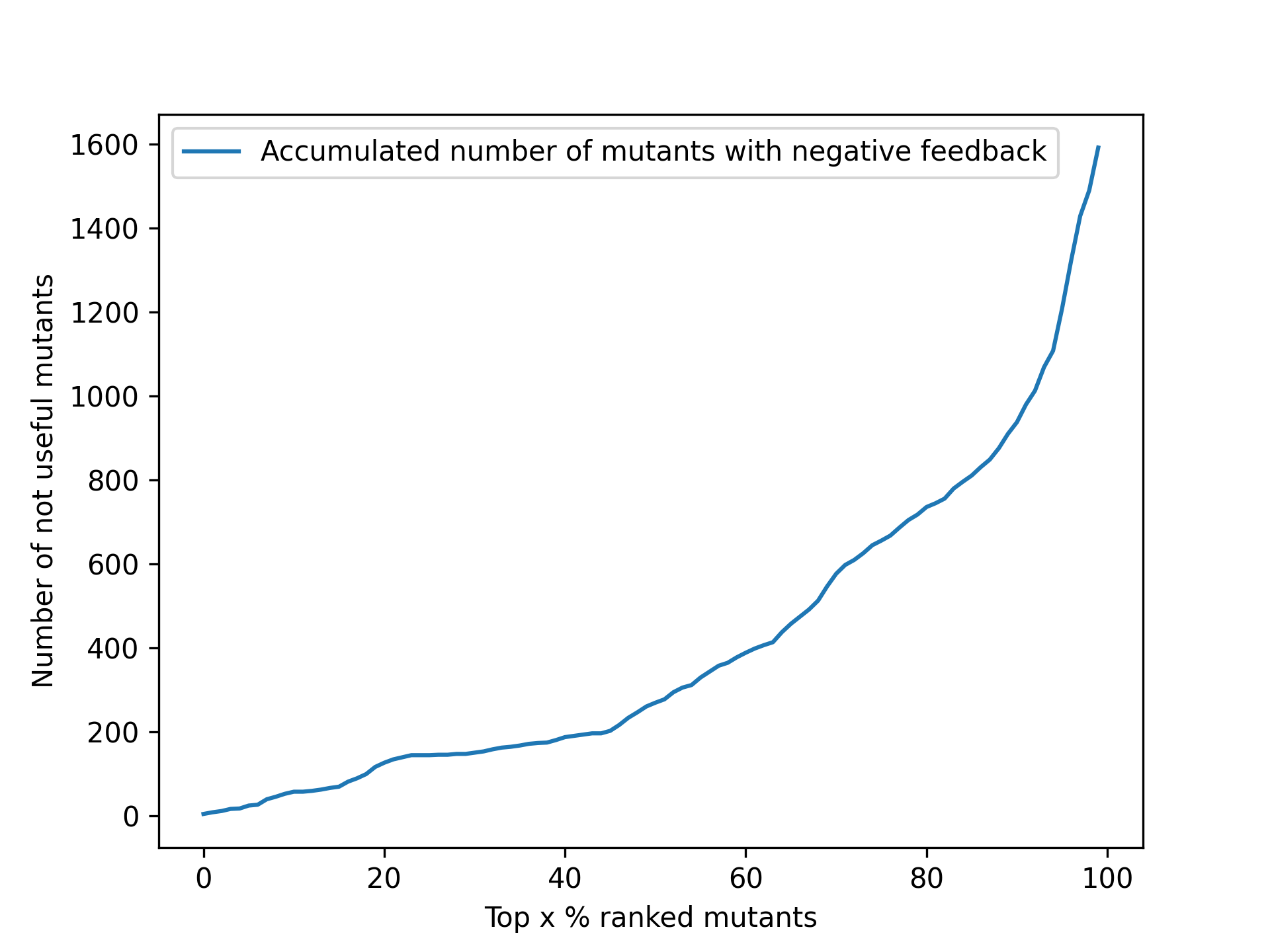}
\caption{Accumulated number of not-useful mutants among the top x \% ranked mutants in the control group.}
\label{fig:rq2_accumulated_number}

\vspace*{6pt}
\includegraphics[width=\linewidth]{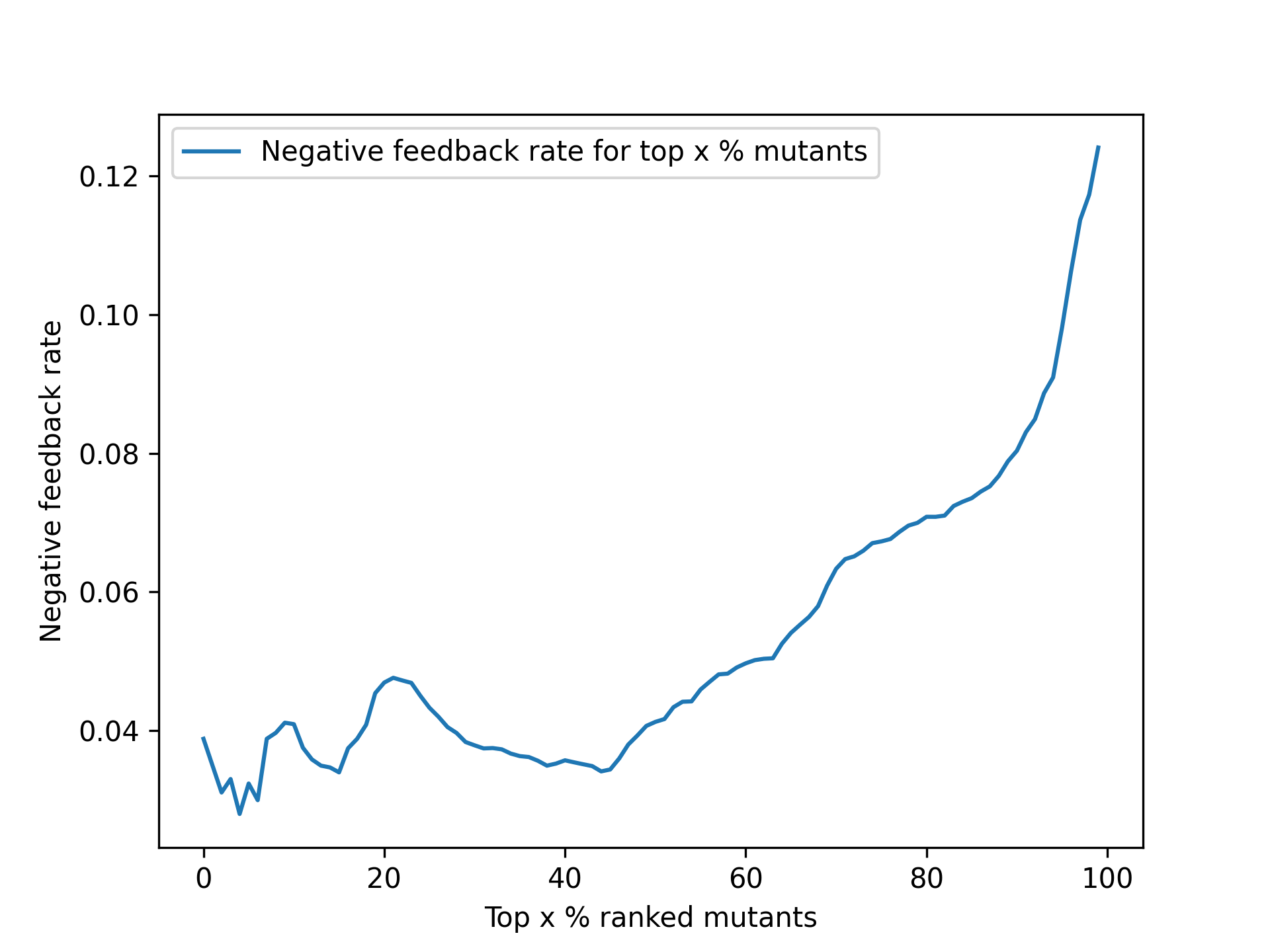}
\caption{Negative feedback rate for top x \% ranked mutants in the control group.}
\label{fig:rq2_not_useful_rate}
\end{figure}

As shown in \autoref{tab:contigency_table}, the control group had \ControlGroupWithFeedback mutants with feedback, and the experiment group had \ExperimentGroupWithFeedback.

For the control group, the rank correlation coefficient (Kendall Tau-b) between \toolname's usefulness score and the developer feedback is \RQtwoControlRankCorrelation ($p < 0.0001$). For the experiment group the correlation coefficient is \RQtwoExperimentRankCorrelation ($p < 0.0001$). These coefficients suggest a weak to moderate correlation and there is no significant difference between the control and experiment groups.

\autoref{fig:rq2_accumulated_number} displays the cumulative number of mutants with negative feedback among the top x\% ranked mutants in the control group. The top 50\% of the ranked mutants only contains \PercentageNotUsefulTopHalf (\NumNotUsefulTopHalf out of \ControlGroupWithNegFeedback) of all mutants that received negative feedback. The negative feedback ratio for the top x\% mutants is shown in \autoref{fig:rq2_not_useful_rate}. It reveals that the negative feedback ratio remains around 4\% for the top half of the ranked mutants, but it steadily increases to 12\% when including all mutants. Both figures suggest that higher ranked mutants are less likely to receive negative feedback.

\autoref{fig:rq2_not_useful_rate} shows a trade-off: while aggressive mutant suppression (e.g., the bottom 40\% of ranked mutants) might lower the negative feedback ratio to about 6\%, it would likely suppress useful mutants as well. We determined how many mutants with positive or negative feedback would have been suppressed in the control group. Since suppression is probabilistic, we computed the expected values:
\begin{itemize}
    \item \NegFeedbackSuppressed out of \ControlGroupWithNegFeedback mutants with negative feedback (\PercentageNegMutantsSuppressed) would have been suppressed. Since these mutants indeed received negative feedback, they are correctly suppressed.

    \item \PosFeedbackSuppressed out of \ControlGroupWithPosFeedback mutants with positive feedback (\PercentagePosMutantsSuppressed) would have been suppressed. Since these mutants received positive feedback, they are incorrectly suppressed.
\end{itemize}

\subsection{RQ3: \rqThree}
\begin{figure}
\centering
\includegraphics[width=\linewidth,keepaspectratio]{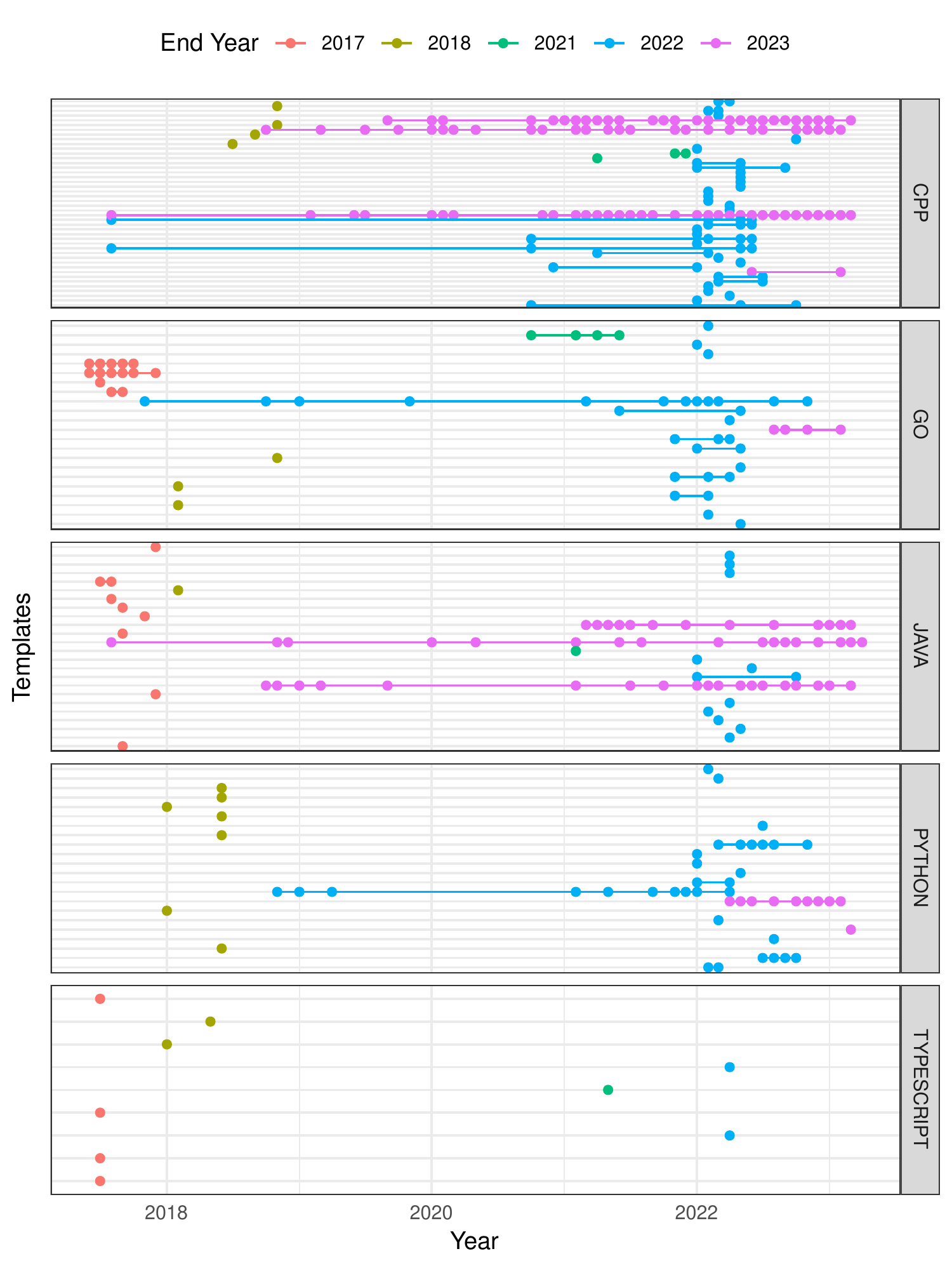}
\caption{Sampled not-useful templates for which at least 50 live mutants surfaced per month on average.}
\label{fig:templates}
\end{figure}
\autoref{fig:templates} shows the results of applying \toolname retroactively to all mutants surfaced since mid 2017 and filtering templates that (1) correspond to at least 50 surfaced mutants per month on average and (2) consistently received negative feedback. Each row in this plot corresponds to a selected template and each dot corresponds to a month during which a template was observable. A line connects multiple months of the same template, indicating the overall time period a template was observable. We manually sampled ten templates for manual inspection across multiple languages and years. For most of them, we were able to identify a corresponding suppression rule in the existing mutation testing service. Below are two concrete examples.

The following GO template corresponds to a common error-handling pattern:
\begin{lstlisting}[language=diff,columns=flexible, frame=tb, basicstyle=\small, breaklines=true]
-if GO_IDENTIFIER_0 != GO_IDENTIFIER_1 { return GO_IDENTIFIER_0 }
\end{lstlisting}
This template corresponds to statement removal mutants that delete \verb|if err != nil { return err }|. Since these statements are ubiquitous but the mutants are not generally worth testing, the negative developer feedback provided a very strong signal for suppression.

The following Java template refers to mutants in annotations:
\begin{lstlisting}[language=diff,columns=flexible, frame=tb, basicstyle=\small, breaklines=true]
-@JAVA_IDENTIFIER_0(JAVA_IDENTIFIER_1 = JAVA_IDENTIFIER_2.JAVA_IDENTIFIER_3, JAVA_IDENTIFIER_4 = true)
+@JAVA_IDENTIFIER_0(JAVA_IDENTIFIER_1 = JAVA_IDENTIFIER_2.JAVA_IDENTIFIER_3, JAVA_IDENTIFIER_4 = !(true))
\end{lstlisting}
This template corresponds to mutants that change a boolean attribute of an annotation from \textit{true} to \textit{!(true)}. Annotations in Java provide metadata or compiler directives, or are used for meta programming. Common examples include the \textit{@Override} annotation, which indicates that a subclass method is overriding a parent class method, the \textit{@Deprecated} annotation, which indicates that a method is deprecated, or the \textit{@Test} annotation for JUnit tests. However, annotations are rarely, if ever, subject to (unit) testing. The clear negative sentiment of developers towards these mutants suggests that suppressing them is arguably the correct choice.

Additionally, we also inspected templates with positive and mixed feedback. Below are two concrete examples.

Among the C++ templates, the mutants corresponding to the following template are considered useful by many developers:
\begin{lstlisting}[language=diff,columns=flexible, frame=tb, basicstyle=\small, breaklines=true]
-} else if (CPP_IDENTIFIER_0 == CPP_IDENTIFIER_1) {
+} else if (true) {
\end{lstlisting}
This template captures mutants that replace a relational operator in the condition of an \textit{else if} statement with the constant value \textit{true}. 

The following Python template has both the highest total PU and PNU count:
\begin{lstlisting}[language=diff,columns=flexible, frame=tb, basicstyle=\small, label={lst:controversial_python_template}, breaklines=true]
-PYTHON_ID_0(PYTHON_ID_1)
\end{lstlisting}
This template captures mutants that remove a function call with a single argument. \citeauthor{just2014mutants} found that statement deletion mutants are often associated with real faults, more so than other types of mutants \cite{just2014mutants}.  Therefore, it is not surprising that many of these mutants are considered useful. At the same time, there are likely many function calls that match this pattern that are not worth testing. One very common class of function call that matches this pattern are telemetry statements, e.g. \texttt{log(id)}. Developers typically do not consider these to be directly impacting user experience and are less interested in testing them specifically. Templates like this are the most controversial:
\begin{align*}
\text{controversy} = PU * (1 - PU) * (PU + PNU)
\end{align*}

 By computing the controversy score for all identifier templates, we found that the most controversial templates for Python, Java, C++, Go, and TypeScript are all statement removal mutants. One of the reasons is that statement removal mutants represent 68\% of all mutants \cite{petrovic2021practical}, which makes them also more likely to receive more feedback. This observation calls for more research to understand the differences between useful statement removal mutants that are associated with real bugs and not-useful statement removal mutants. For example, future research can inform additional context that should be considered to distinguish these.
 
 \subsection{Discussion}
It is important to put the results in perspective when assessing their practical significance. First, we are comparing \toolname against a strong baseline, which employs many suppression heuristics that have lowered the negative feedback ratio from over 80\%, during initial development, to well below 15\% now. Second, the negative feedback ratio is a metric derived from developer feedback, which is subjective. A mutant can receive negative feedback for many reasons---some mutants represent infeasible test goals, others are simply not worth satisfying~\cite{offutt1997automatically,petrovic2018industrial}. While many not-useful mutants consistently receive negative feedback, others receive no or even inconclusive feedback. For example, the very same mutant can receive different feedback from the author and reviewers during a code review. As a result, the lower bound for the negative feedback ratio is very likely greater than zero. Since our baseline is already approaching a ratio in the single digits, it is important to keep in mind that the true lower bound, and thus the best possible result, could be anywhere between 0\% and 10\%.

\toolname suppresses about 50\% mutants with negative feedback, at the cost of suppressing about \PercentagePosMutantsSuppressed  mutants with positive feedback. One of the reasons why \toolname does not suppress more not-useful mutants is that the corresponding identifier templates have not yet received enough negative feedback, and as a result the suppression function conservatively retains them. Indeed, we have observed that about 80\% of all templates correspond to at most ten surfaced mutants---that is, the templates are too specific. At the same time, the current templates do not consider enough context to retain all useful mutants and to disambiguate mutants with mixed feedback---that is, the templates are too generic. We leave a deeper investigation and possible refinements for future work.

Suppressing mutants can also be achieved through the implementation of static rules that prevent mutants from being generated in the first place, e.g., mutations of logging statements. The existing mutation testing service already implements this using arid nodes, which are AST nodes that are identified through pattern matching and that are never mutated. While effective, \toolname offers several advantages over the arid node approach. It automatically creates an identifier template that is used to suppress a particular type of mutant, eliminating the need for manual implementation of static rules. In fact, we have discovered multiple identifier templates that correspond to our existing static rules, demonstrating that \toolname is able to recover these rules. Overall, the combination of using arid-node heuristics and \toolname provides an effective strategy for suppressing mutants in a mutation testing service.
\section{Related Work}
\label{sec:related}

The closest related work in terms of industry-scale mutation testing is \citeauthor{beller2021would}'s study of mutation testing at Facebook. They also focused on diff-based mutation testing, but a key difference is that their approach uses semi-automatic learning on common Java bug patterns to create a small number of mutation operators that yield mostly useful mutants by design. In contrast, our approach uses a predefined set of mutation operators and suppresses not-useful mutants. The two approaches could potentially be combined for greater effect.

\subsection{Mutant Selection and Prioritization}
A number of studies have shown that a small percentage of mutants are sufficient, that there is substantial redundancy among generated mutants, and that many generated mutants are not useful~\cite{petrovic2021does,ChenGTEHFAJ2020,kurtz2016analyzing}. We refer the reader to the literature review by \citeauthor{pizzoleto2019systematic} for more comprehensive related work, especially PG-1 (\textit{Reducing the number of mutants}) and PG-5 (\textit{Avoiding the creation of certain mutants}) mentioned in the paper for the closest related work regarding mutation selection or prioritization.

\citeauthor{gopinath2017mutation} compared multiple strategies for reducing the number of mutants against random sampling \cite{gopinath2017mutation}. They found that none of the strategies yield an effective advantage larger than 5\%, when compared with random sampling. Given the result, they warn against adopting mutant reduction techniques without adequate reason. In an industrial context, however, a 1\% reduction in mutants with negative feedback could be worthwhile (it safes valuable developer time) as long as mutants with positive feedback are retained.

\citeauthor{brown2017care} introduced an approach that extracts mutation operators from the revision history of software projects~\cite{brown2017care}. The approach is based on the observation that if a commit corrects a bug, then its reversal is potentially introducing a bug, making it a viable mutation operator. Using this approach, they mined mutation operators from the revision history of the top-50 most forked C projects on GitHub. Their findings showed that mutants generated from the mined mutation operators were just as challenging to kill as mutants generated using traditional mutation operators. However, the mined mutation operators exhibited greater diversity in the types of changes than traditional mutation operators.

\subsection{Mutant Usefulness}

\citeauthor{JustKA2017b} studied the degree to which program context, defined over the abstract syntax tree, is correlated with a mutant's expected usefulness. In this work, usefulness was termed mutant utility and measured along three dimensions: equivalence, triviality, and dominance. The results showed that mutant utility is context-dependent: the same mutation may lead to a high-utility mutant in one context but not not necessarily in another.

\citeauthor{kaufman2022prioritizing} proposed test completeness advancement probability (TCAP) as a measure to define how useful a mutant is \cite{kaufman2022prioritizing}. TCAP is a probability that if a mutant is presented as a test goal, it will elicit a test to improve the test completeness. Evaluated on 9 projects from the Defects4J benchmark~\cite{just2014defects4j}, they concluded that TCAP can be predicted from program context and that TCAP-based mutant prioritization improves test completeness faster than the previous state-of-the-art, which is random prioritization.

\toolname and its evaluation differ from prior work in two key aspects. First, \toolname uses an abstraction over the unified diff, which encodes both the mutation and the surrounding context. Second, mutant usefulness is derived from developer feedback as opposed to test or mutant characteristics.

\subsection{Diff-based Mutation Testing}

\citeauthor{cachia2013towards} focused on industrial adoption of mutation testing, and suggested incremental mutation testing that limits the scope of mutation testing to the changed code \cite{cachia2013towards}. In their evaluation, they found that incremental mutation testing reduced the number of generated mutants and the execution time of mutation testing.

\citeauthor{ma2020commit} studied the relationship between commit-aware mutation testing and traditional mutation testing \cite{ma2020commit}. They found that the commit-aware mutation score and the traditional mutation score are only weakly correlated, and that mutants from traditional mutation testing have a 30\% lower chance of revealing faults introduced by the commit.
\section{Conclusions}
\label{sec:concl}

This paper resents \toolname, an approach designed to enhance the user experience of mutation testing by reducing the negative feedback ratio of mutants surfaced during code review. \toolname ranks an suppresses newly generated mutants based on identifier templates that group similar mutants and aggregate historical developer feedback for said mutants.

Based on an A/B testing study, we found that the negative feedback ratio for \toolname (\ExperimentGroupNegativeRate) is statically significantly lower, compared to manually tuned baseline. The results also show that mutants higher ranked by \toolname have a lower probability of receiving negative feedback. Additionally, we observe that \toolname would have suppressed about 50\% of not-useful mutants but also \PercentagePosMutantsSuppressed of useful mutants in the control group, highlighting a trade-off. Finally, we identified statement deletion mutants as the type of mutants whose identifier templates received both a high number of positive and a high number of negative feedback. Going forward, we aim to investigate what distinguishes useful statement deletion mutants from not-useful ones despite their similarities w.r.t. identifier templates. Furthermore, we would like to understand whether \toolname should apply different abstractions for the mutated lines vs. context lines.

One important result of \toolname in an industrial setting is its lower cost of development and maintenance, compared to manually implementing suppression rules. Our results provide evidence that \toolname produces identifier templates that correspond to static suppression rules implemented in the past, but future work should further evaluate to what extent \toolname can fully replace suppression rules or to what extent the two approaches are complementary.

\balance
\bibliography{references}


\begin{thebibliography}{00}


\ifx \showCODEN    \undefined \def \showCODEN     #1{\unskip}     \fi
\ifx \showDOI      \undefined \def \showDOI       #1{#1}\fi
\ifx \showISBNx    \undefined \def \showISBNx     #1{\unskip}     \fi
\ifx \showISBNxiii \undefined \def \showISBNxiii  #1{\unskip}     \fi
\ifx \showISSN     \undefined \def \showISSN      #1{\unskip}     \fi
\ifx \showLCCN     \undefined \def \showLCCN      #1{\unskip}     \fi
\ifx \shownote     \undefined \def \shownote      #1{#1}          \fi
\ifx \showarticletitle \undefined \def \showarticletitle #1{#1}   \fi
\ifx \showURL      \undefined \def \showURL       {\relax}        \fi
\providecommand\bibfield[2]{#2}
\providecommand\bibinfo[2]{#2}
\providecommand\natexlab[1]{#1}
\providecommand\showeprint[2][]{arXiv:#2}

\bibitem[\protect\citeauthoryear{Acree, Budd, DeMillo, Lipton, and
  Sayward}{Acree et~al\mbox{.}}{1979}]%
        {acree1979mutation}
\bibfield{author}{\bibinfo{person}{Allen~T Acree}, \bibinfo{person}{Timothy~A
  Budd}, \bibinfo{person}{Richard~A DeMillo}, \bibinfo{person}{Richard~J
  Lipton}, {and} \bibinfo{person}{Frederick~G Sayward}.}
  \bibinfo{year}{1979}\natexlab{}.
\newblock \bibinfo{booktitle}{{\em Mutation Analysis.}}
\newblock \bibinfo{type}{{T}echnical {R}eport}. \bibinfo{institution}{Georgia
  Inst of Tech Atlanta School of Information And Computer Science}.
\newblock


\bibitem[\protect\citeauthoryear{Beller, Wong, Bader, Scott, Machalica,
  Chandra, and Meijer}{Beller et~al\mbox{.}}{2021}]%
        {beller2021would}
\bibfield{author}{\bibinfo{person}{Moritz Beller}, \bibinfo{person}{Chu-Pan
  Wong}, \bibinfo{person}{Johannes Bader}, \bibinfo{person}{Andrew Scott},
  \bibinfo{person}{Mateusz Machalica}, \bibinfo{person}{Satish Chandra}, {and}
  \bibinfo{person}{Erik Meijer}.} \bibinfo{year}{2021}\natexlab{}.
\newblock \showarticletitle{What it would take to use mutation testing in
  industry—a study at facebook}. In \bibinfo{booktitle}{{\em International
  Conference on Software Engineering: Software Engineering in Practice
  (ICSE-SEIP)}}. IEEE, \bibinfo{pages}{268--277}.
\newblock


\bibitem[\protect\citeauthoryear{Brown, Vaughn, Liblit, and Reps}{Brown
  et~al\mbox{.}}{2017}]%
        {brown2017care}
\bibfield{author}{\bibinfo{person}{David~Bingham Brown},
  \bibinfo{person}{Michael Vaughn}, \bibinfo{person}{Ben Liblit}, {and}
  \bibinfo{person}{Thomas Reps}.} \bibinfo{year}{2017}\natexlab{}.
\newblock \showarticletitle{The Care and Feeding of Wild-Caught Mutants}. In
  \bibinfo{booktitle}{{\em Proceedings of the Joint Meeting of the European
  Software Engineering Conference and the Symposium on the Foundations of
  Software Engineering (ESEC/FSE)}} {\em (\bibinfo{series}{ESEC/FSE 2017})}.
  \bibinfo{publisher}{Association for Computing Machinery},
  \bibinfo{address}{New York, NY, USA}, \bibinfo{pages}{511–522}.
\newblock
\showISBNx{9781450351058}
\showDOI{%
\url{https://doi.org/10.1145/3106237.3106280}}


\bibitem[\protect\citeauthoryear{Cachia, Micallef, and Colombo}{Cachia
  et~al\mbox{.}}{2013}]%
        {cachia2013towards}
\bibfield{author}{\bibinfo{person}{Mark~Anthony Cachia}, \bibinfo{person}{Mark
  Micallef}, {and} \bibinfo{person}{Christian Colombo}.}
  \bibinfo{year}{2013}\natexlab{}.
\newblock \showarticletitle{Towards incremental mutation testing}.
\newblock \bibinfo{journal}{{\em Electronic Notes in Theoretical Computer
  Science (ENTCS)\/}}  \bibinfo{volume}{294} (\bibinfo{year}{2013}),
  \bibinfo{pages}{2--11}.
\newblock


\bibitem[\protect\citeauthoryear{Chen, Gopinath, Tadakamalla, Ernst, Holmes,
  Fraser, Ammann, and Just}{Chen et~al\mbox{.}}{2020}]%
        {ChenGTEHFAJ2020}
\bibfield{author}{\bibinfo{person}{Yiqun~T. Chen}, \bibinfo{person}{Rahul
  Gopinath}, \bibinfo{person}{Anita Tadakamalla}, \bibinfo{person}{Michael~D.
  Ernst}, \bibinfo{person}{Reid Holmes}, \bibinfo{person}{Gordon Fraser},
  \bibinfo{person}{Paul Ammann}, {and} \bibinfo{person}{Ren{\'e} Just}.}
  \bibinfo{year}{2020}\natexlab{}.
\newblock \showarticletitle{Revisiting the Relationship Between Fault
  Detection, Test Adequacy Criteria, and Test Set Size}. In
  \bibinfo{booktitle}{{\em Proceedings of the International Conference on
  Automated Software Engineering (ASE)}}. \bibinfo{pages}{237--249}.
\newblock


\bibitem[\protect\citeauthoryear{Gopinath, Ahmed, Alipour, Jensen, and
  Groce}{Gopinath et~al\mbox{.}}{2017}]%
        {gopinath2017mutation}
\bibfield{author}{\bibinfo{person}{Rahul Gopinath}, \bibinfo{person}{Iftekhar
  Ahmed}, \bibinfo{person}{Mohammad~Amin Alipour}, \bibinfo{person}{Carlos
  Jensen}, {and} \bibinfo{person}{Alex Groce}.}
  \bibinfo{year}{2017}\natexlab{}.
\newblock \showarticletitle{Mutation reduction strategies considered harmful}.
\newblock \bibinfo{journal}{{\em IEEE Transactions on Reliability\/}}
  \bibinfo{volume}{66}, \bibinfo{number}{3} (\bibinfo{year}{2017}),
  \bibinfo{pages}{854--874}.
\newblock


\bibitem[\protect\citeauthoryear{Jia and Harman}{Jia and Harman}{2010}]%
        {jia2010analysis}
\bibfield{author}{\bibinfo{person}{Yue Jia} {and} \bibinfo{person}{Mark
  Harman}.} \bibinfo{year}{2010}\natexlab{}.
\newblock \showarticletitle{An analysis and survey of the development of
  mutation testing}.
\newblock \bibinfo{journal}{{\em IEEE Transactions on Software Engineering
  (TSE)\/}} \bibinfo{volume}{37}, \bibinfo{number}{5} (\bibinfo{year}{2010}),
  \bibinfo{pages}{649--678}.
\newblock


\bibitem[\protect\citeauthoryear{Just, Jalali, and Ernst}{Just
  et~al\mbox{.}}{2014a}]%
        {just2014defects4j}
\bibfield{author}{\bibinfo{person}{Ren{\'e} Just}, \bibinfo{person}{Darioush
  Jalali}, {and} \bibinfo{person}{Michael~D Ernst}.}
  \bibinfo{year}{2014}\natexlab{a}.
\newblock \showarticletitle{Defects4J: A database of existing faults to enable
  controlled testing studies for Java programs}. In \bibinfo{booktitle}{{\em
  Proceedings of the International Symposium on Software Testing and Analysis
  (ISSTA)}}. \bibinfo{pages}{437--440}.
\newblock


\bibitem[\protect\citeauthoryear{Just, Jalali, Inozemtseva, Ernst, Holmes, and
  Fraser}{Just et~al\mbox{.}}{2014b}]%
        {just2014mutants}
\bibfield{author}{\bibinfo{person}{Ren{\'e} Just}, \bibinfo{person}{Darioush
  Jalali}, \bibinfo{person}{Laura Inozemtseva}, \bibinfo{person}{Michael~D
  Ernst}, \bibinfo{person}{Reid Holmes}, {and} \bibinfo{person}{Gordon
  Fraser}.} \bibinfo{year}{2014}\natexlab{b}.
\newblock \showarticletitle{Are mutants a valid substitute for real faults in
  software testing?}. In \bibinfo{booktitle}{{\em Proceedings of the Symposium
  on the Foundations of Software Engineering (FSE)}}.
  \bibinfo{pages}{654--665}.
\newblock


\bibitem[\protect\citeauthoryear{Just, Kurtz, and Ammann}{Just
  et~al\mbox{.}}{2017}]%
        {JustKA2017b}
\bibfield{author}{\bibinfo{person}{Ren{\'e} Just}, \bibinfo{person}{Bob Kurtz},
  {and} \bibinfo{person}{Paul Ammann}.} \bibinfo{year}{2017}\natexlab{}.
\newblock \showarticletitle{Inferring Mutant Utility from Program Context}. In
  \bibinfo{booktitle}{{\em Proceedings of the International Symposium on
  Software Testing and Analysis (ISSTA)}}. \bibinfo{pages}{284--294}.
\newblock


\bibitem[\protect\citeauthoryear{Kaufman, Featherman, Alvin, Kurtz, Ammann, and
  Just}{Kaufman et~al\mbox{.}}{2022}]%
        {kaufman2022prioritizing}
\bibfield{author}{\bibinfo{person}{Samuel~J Kaufman}, \bibinfo{person}{Ryan
  Featherman}, \bibinfo{person}{Justin Alvin}, \bibinfo{person}{Bob Kurtz},
  \bibinfo{person}{Paul Ammann}, {and} \bibinfo{person}{Ren{\'e} Just}.}
  \bibinfo{year}{2022}\natexlab{}.
\newblock \showarticletitle{Prioritizing mutants to guide mutation testing}. In
  \bibinfo{booktitle}{{\em Proceedings of the International Conference on
  Software Engineering (ICSE)}}. \bibinfo{pages}{1743--1754}.
\newblock


\bibitem[\protect\citeauthoryear{Kochhar, Thung, and Lo}{Kochhar
  et~al\mbox{.}}{2015}]%
        {kochhar2015code}
\bibfield{author}{\bibinfo{person}{Pavneet~Singh Kochhar},
  \bibinfo{person}{Ferdian Thung}, {and} \bibinfo{person}{David Lo}.}
  \bibinfo{year}{2015}\natexlab{}.
\newblock \showarticletitle{Code coverage and test suite effectiveness:
  Empirical study with real bugs in large systems}. In \bibinfo{booktitle}{{\em
  IEEE International Conference on Software Analysis, Evolution and
  Reengineering (SANER)}}. IEEE, \bibinfo{pages}{560--564}.
\newblock


\bibitem[\protect\citeauthoryear{Kurtz, Ammann, Offutt, Delamaro, Kurtz, and
  G{\"o}k{\c{c}}e}{Kurtz et~al\mbox{.}}{2016}]%
        {kurtz2016analyzing}
\bibfield{author}{\bibinfo{person}{Bob Kurtz}, \bibinfo{person}{Paul Ammann},
  \bibinfo{person}{Jeff Offutt}, \bibinfo{person}{M{\'a}rcio~E Delamaro},
  \bibinfo{person}{Mariet Kurtz}, {and} \bibinfo{person}{Nida
  G{\"o}k{\c{c}}e}.} \bibinfo{year}{2016}\natexlab{}.
\newblock \showarticletitle{Analyzing the validity of selective mutation with
  dominator mutants}. In \bibinfo{booktitle}{{\em Proceedings of the Symposium
  on the Foundations of Software Engineering (FSE)}}.
  \bibinfo{pages}{571--582}.
\newblock


\bibitem[\protect\citeauthoryear{Ma, Laurent, Ojdani{\'c}, Chekam, Ventresque,
  and Papadakis}{Ma et~al\mbox{.}}{2020}]%
        {ma2020commit}
\bibfield{author}{\bibinfo{person}{Wei Ma}, \bibinfo{person}{Thomas Laurent},
  \bibinfo{person}{Milo{\v{s}} Ojdani{\'c}}, \bibinfo{person}{Thierry~Titcheu
  Chekam}, \bibinfo{person}{Anthony Ventresque}, {and} \bibinfo{person}{Mike
  Papadakis}.} \bibinfo{year}{2020}\natexlab{}.
\newblock \showarticletitle{Commit-aware mutation testing}. In
  \bibinfo{booktitle}{{\em International Conference on Software Maintenance and
  Evolution}}. IEEE, \bibinfo{pages}{394--405}.
\newblock


\bibitem[\protect\citeauthoryear{Offutt and Pan}{Offutt and Pan}{1997}]%
        {offutt1997automatically}
\bibfield{author}{\bibinfo{person}{A~Jefferson Offutt} {and}
  \bibinfo{person}{Jie Pan}.} \bibinfo{year}{1997}\natexlab{}.
\newblock \showarticletitle{Automatically detecting equivalent mutants and
  infeasible paths}.
\newblock \bibinfo{journal}{{\em Software Testing, Verification and Reliability
  (JSTVR)\/}} \bibinfo{volume}{7}, \bibinfo{number}{3} (\bibinfo{year}{1997}),
  \bibinfo{pages}{165--192}.
\newblock


\bibitem[\protect\citeauthoryear{Pearson}{Pearson}{1900}]%
        {pearson1900x}
\bibfield{author}{\bibinfo{person}{Karl Pearson}.}
  \bibinfo{year}{1900}\natexlab{}.
\newblock \showarticletitle{X. On the criterion that a given system of
  deviations from the probable in the case of a correlated system of variables
  is such that it can be reasonably supposed to have arisen from random
  sampling}.
\newblock \bibinfo{journal}{{\em The London, Edinburgh, and Dublin
  Philosophical Magazine and Journal of Science\/}} \bibinfo{volume}{50},
  \bibinfo{number}{302} (\bibinfo{year}{1900}), \bibinfo{pages}{157--175}.
\newblock


\bibitem[\protect\citeauthoryear{Petrovi{\'c}, Ivankovi{\'c}, Fraser, and
  Just}{Petrovi{\'c} et~al\mbox{.}}{2021}]%
        {petrovic2021does}
\bibfield{author}{\bibinfo{person}{Goran Petrovi{\'c}}, \bibinfo{person}{Marko
  Ivankovi{\'c}}, \bibinfo{person}{Gordon Fraser}, {and}
  \bibinfo{person}{Ren{\'e} Just}.} \bibinfo{year}{2021}\natexlab{}.
\newblock \showarticletitle{Does mutation testing improve testing practices?}.
  In \bibinfo{booktitle}{{\em Proceedings of the International Conference on
  Software Engineering (ICSE)}}. \bibinfo{pages}{910--921}.
\newblock


\bibitem[\protect\citeauthoryear{Petrovic, Ivankovic, Fraser, and
  Just}{Petrovic et~al\mbox{.}}{2021}]%
        {petrovic2021practical}
\bibfield{author}{\bibinfo{person}{Goran Petrovic}, \bibinfo{person}{Marko
  Ivankovic}, \bibinfo{person}{Gordon Fraser}, {and} \bibinfo{person}{Ren{\'e}
  Just}.} \bibinfo{year}{2021}\natexlab{}.
\newblock \showarticletitle{Practical mutation testing at scale: A view from
  Google}.
\newblock \bibinfo{journal}{{\em IEEE Transactions on Software Engineering
  (TSE)\/}} (\bibinfo{year}{2021}).
\newblock


\bibitem[\protect\citeauthoryear{Petrovic, Ivankovic, Kurtz, Ammann, and
  Just}{Petrovic et~al\mbox{.}}{2018}]%
        {petrovic2018industrial}
\bibfield{author}{\bibinfo{person}{Goran Petrovic}, \bibinfo{person}{Marko
  Ivankovic}, \bibinfo{person}{Bob Kurtz}, \bibinfo{person}{Paul Ammann}, {and}
  \bibinfo{person}{Ren{\'e} Just}.} \bibinfo{year}{2018}\natexlab{}.
\newblock \showarticletitle{An industrial application of mutation testing:
  Lessons, challenges, and research directions}. In \bibinfo{booktitle}{{\em
  Proceedings of the International Conference on Software Testing, Verification
  and Validation Workshops (ICSTW)}}. \bibinfo{pages}{47--53}.
\newblock


\bibitem[\protect\citeauthoryear{Pizzoleto, Ferrari, Offutt, Fernandes, and
  Ribeiro}{Pizzoleto et~al\mbox{.}}{2019}]%
        {pizzoleto2019systematic}
\bibfield{author}{\bibinfo{person}{Alessandro~Viola Pizzoleto},
  \bibinfo{person}{Fabiano~Cutigi Ferrari}, \bibinfo{person}{Jeff Offutt},
  \bibinfo{person}{Leo Fernandes}, {and} \bibinfo{person}{M{\'a}rcio Ribeiro}.}
  \bibinfo{year}{2019}\natexlab{}.
\newblock \showarticletitle{A systematic literature review of techniques and
  metrics to reduce the cost of mutation testing}.
\newblock \bibinfo{journal}{{\em Journal of Systems and Software (JSS)\/}}
  \bibinfo{volume}{157} (\bibinfo{year}{2019}), \bibinfo{pages}{110388}.
\newblock


\bibitem[\protect\citeauthoryear{Sadowski, Aftandilian, Eagle, Miller-Cushon,
  and Jaspan}{Sadowski et~al\mbox{.}}{2018}]%
        {sadowski2018lessons}
\bibfield{author}{\bibinfo{person}{Caitlin Sadowski}, \bibinfo{person}{Edward
  Aftandilian}, \bibinfo{person}{Alex Eagle}, \bibinfo{person}{Liam
  Miller-Cushon}, {and} \bibinfo{person}{Ciera Jaspan}.}
  \bibinfo{year}{2018}\natexlab{}.
\newblock \showarticletitle{Lessons from building static analysis tools at
  {G}oogle}.
\newblock \bibinfo{journal}{{\em Communications of the ACM (CACM)\/}}
  \bibinfo{volume}{61}, \bibinfo{number}{4} (\bibinfo{year}{2018}),
  \bibinfo{pages}{58--66}.
\newblock


\bibitem[\protect\citeauthoryear{Wong and Mathur}{Wong and Mathur}{1995}]%
        {wong1995reducing}
\bibfield{author}{\bibinfo{person}{W~Eric Wong} {and} \bibinfo{person}{Aditya~P
  Mathur}.} \bibinfo{year}{1995}\natexlab{}.
\newblock \showarticletitle{Reducing the cost of mutation testing: An empirical
  study}.
\newblock \bibinfo{journal}{{\em Journal of Systems and Software (JSS)\/}}
  \bibinfo{volume}{31}, \bibinfo{number}{3} (\bibinfo{year}{1995}),
  \bibinfo{pages}{185--196}.
\newblock


\end{thebibliography}
\bibliographystyle{ACM-Reference-Format}

\end{document}